\documentclass[conference]{IEEEtran}
\IEEEoverridecommandlockouts

\usepackage{cite}
\usepackage{amsmath,amssymb,amsfonts}
\usepackage{algorithmic}
\usepackage{graphicx}
\usepackage{textcomp}
\usepackage{xcolor}
\usepackage{float} 
\usepackage{listings}
\usepackage[T1]{fontenc}
\usepackage[utf8]{inputenc}
\usepackage{url}
\def\BibTeX{{\rm B\kern-.05em{\sc i\kern-.025em b}\kern-.08em
    T\kern-.1667em\lower.7ex\hbox{E}\kern-.125emX}}
\begin{document}

\title{PHICOIN (PHI): The Proof of Work High-Performance Infrastructure}

\author{
\IEEEauthorblockN{Guang Yang\IEEEauthorrefmark{1}, Peter Trinh\IEEEauthorrefmark{1}, Sannan Iqbal\IEEEauthorrefmark{1}, Justin Zhang\IEEEauthorrefmark{2}}
\IEEEauthorblockA{\IEEEauthorrefmark{1}University of California, Berkeley\\
Email: \{guangyang19, trinhp, siqbal\}@berkeley.edu}
\IEEEauthorblockA{\IEEEauthorrefmark{2}Yale University\\
Email: justin.zhang.jz932@yale.edu}
}

\maketitle

\begin{abstract}
    PHICOIN (PHI) is a high-performance cryptocurrency based on the Proof-of-Work (PoW) mechanism. It aims to provide ordinary users with decentralized participation opportunities through an improved and innovative mining algorithm and fair design principles. PHI addresses the challenges of centralization in cryptocurrency mining by enhancing resistance to ASIC and FPGA devices and promoting fair participation. This paper outlines the technical specifications, mission, and roadmap for PHI, highlighting its potential to become a foundational infrastructure for PoW cryptocurrencies.
\end{abstract}


\section{Introduction}

In the early days of Bitcoin (BTC), the Proof-of-Work (PoW) mechanism provided ordinary people with a fair and decentralized opportunity to participate \cite{nakamoto2008bitcoin}. However, as the mining industry evolved, the widespread use of FPGA and ASIC devices turned mining into a centralized business, making it increasingly difficult for average individuals to profit from PoW mining.

BTC mining has become more specialized and centralized, raising the barrier to entry and hindering the essence of decentralization. Additionally, Bitcoin's low transactions per second (TPS) and lack of effective scalability make it insufficient for modern financial applications.

Since Ethereum (ETH) transitioned to the Proof-of-Stake (PoS) mechanism \cite{buterin2014next}, many traditional GPU miners have lost stable mining income. Specialized ASIC devices have dominated the traditional Ethash mining market, and large-scale professional mining farms use outdated graphics cards (like RX470). These GPUs, large in scale and centralized like ASICs, create obstacles for ordinary people to participate, further impeding decentralization.

Moreover, large miners with access to cheap electricity resources have further monopolized the mining market, leading to centralization among miners and transactions, leaving ordinary users with virtually no chance to participate.

These issues have resulted in the market lacking a truly fair, decentralized, high-performance, and scalable PoW cryptocurrency. To address these long-standing pain points, we introduce PHICOIN—the PoW High-Performance Infrastructure cryptocurrency, aiming to become a foundational infrastructure for PoW cryptocurrencies. 

We envision PHICOIN as a PoW cryptocurrency with high performance, capable of swiftly phasing out outdated equipment while maintaining continuous mining opportunities. It offers fair participation to everyone, allowing individuals to use their own equipment to contribute collectively and maintain the network. This project will belong to all participants, emphasizing complete decentralization and ensuring that the infrastructure is maintained by everyone.

\section{Solution}

PHI is a high-performance cryptocurrency based on the PoW mechanism. It aims to provide ordinary users with decentralized participation opportunities through an improved and innovative mining algorithm and fair design principles.

PHI's mining algorithm, \textbf{Phihash}, is based on the Ethash cache structure and incorporates the randomness of the KawPow/ProgPow algorithms \cite{progpow}. It uses the Permuted Congruential Generator (PCG) technology to increase the unpredictability of branches. Additionally, we employ lookup table technology to ensure accurate branch hits, further enhancing the algorithm's randomness and resisting replication attacks from ASIC and FPGA devices.
\begin{figure}[H]
    \centering
    \includegraphics[width=0.8\linewidth]{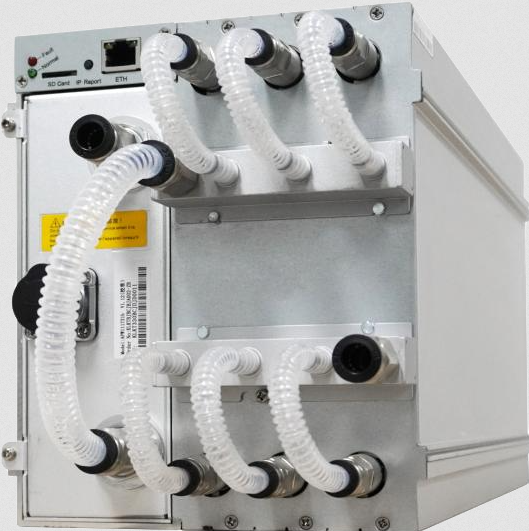} 
    \caption{ Bitmain Antminer S21 XP Hydro 473.00 Th/s @ 5676W  \cite{whattomine2024}}
    \label{fig:pcg32_vs_kiss99}
\end{figure}
An FPGA (Field-Programmable Gate Array) is an integrated circuit that can be programmed after manufacturing to perform specific tasks efficiently, offering flexibility but requiring specialized knowledge to configure. An ASIC (Application-Specific Integrated Circuit), on the other hand, is a chip designed for a specific purpose, such as cryptocurrency mining, delivering superior performance at the expense of flexibility.
\begin{figure}[H]
    \centering
    \includegraphics[width=1\linewidth]{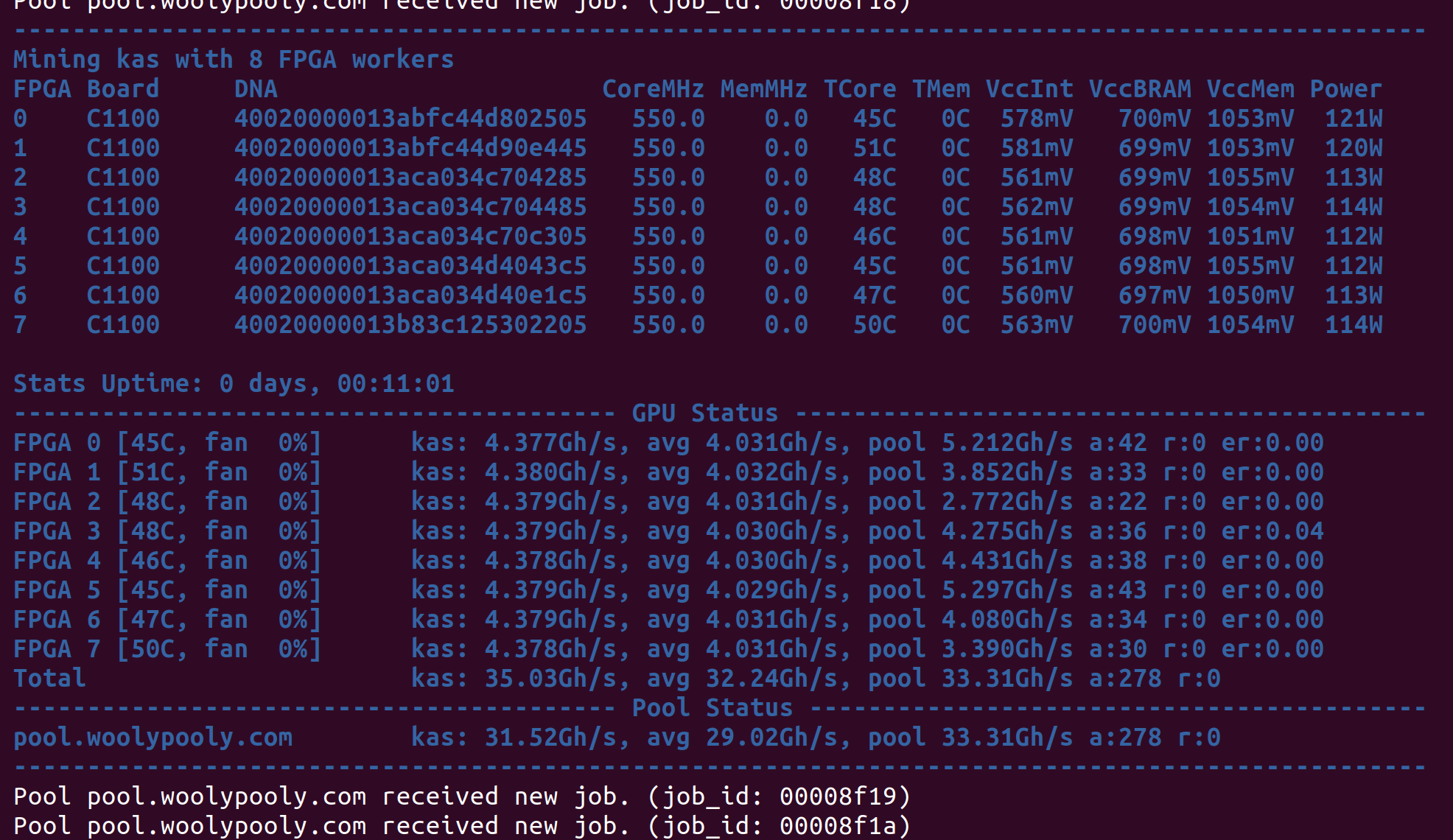} 
    \caption{ Xilinx C1100 mining ETHash/ETChash at 4.37Gh/s \cite{hardforum} }
    \label{fig:pcg32_vs_kiss99}
\end{figure}
The use of FPGA and ASIC devices contributes to the centralization of mining due to several factors. First, the technical barriers to entry are high, as developing and optimizing FPGA configurations or designing ASIC chips requires expertise in hardware programming, chip design, and algorithm tuning. \cite{xilinx_C1100} Second, development and operational costs are significant, with ASIC manufacturing involving high upfront capital investment, making it accessible primarily to large organizations. Finally, maintenance and infrastructure requirements—such as specialized cooling systems, power management, and data center facilities—further restrict participation to professional operators with access to industrial-grade setups.

This centralization poses a threat to the decentralized nature of blockchain networks, as large-scale mining farms can dominate the hashrate by leveraging optimized hardware and access to cheap electricity \cite{whattomine2024}. Such concentration of computational power increases the risk of network control, undermining both security and fairness. PHI aims to mitigate these risks by designing an algorithm that remains accessible to ordinary users with consumer-grade equipment, ensuring equal opportunities for participation and long-term network decentralization.

To achieve fairer participation, we have increased the memory requirement over 4\,GB, ensuring that only users with modern GPUs can participate in mining. This approach not only prevents participation from outdated GPUs held by large miners—which are power-hungry, large-scale, and highly centralized, causing resource monopolization—but also effectively reduces their impact on the mining process. \cite{StatistaSteam2024}

\begin{figure}[H]
    \centering
    \includegraphics[width=1\linewidth]{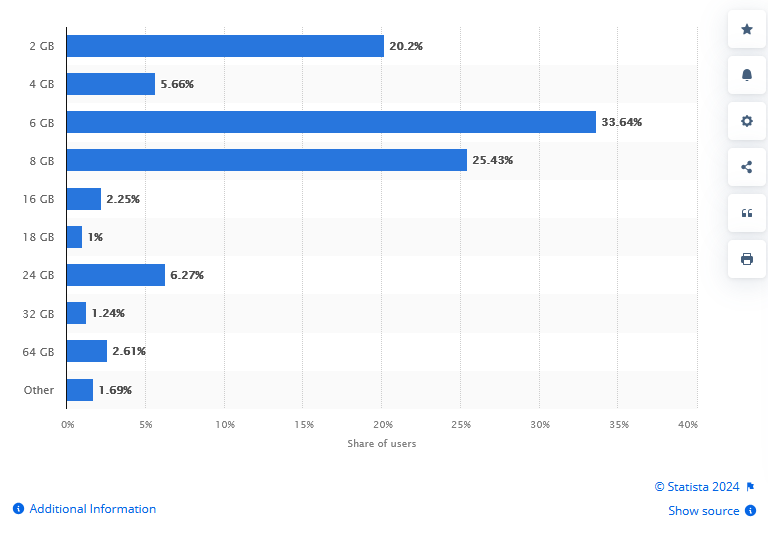} 
    \caption{ Video random-access memory (VRAM) share among Steam users as of March 2024 }
    \label{fig:pcg32_vs_kiss99}
\end{figure}

We have also added FP32 logic computation branches in the mining algorithm, fully utilizing the high floating-point computing capabilities of GPUs. FP32 arithmetic has long been the backbone of GPU performance, playing a crucial role in applications from gaming to deep learning. GPUs excel at high-throughput FP32 computations by leveraging their parallel architectures, with thousands of cores capable of simultaneously performing floating-point operations. For example, Nvidia's Pascal architecture (2016) delivered 10.6 TFLOPS of FP32 performance, while more recent architectures like Ampere have pushed FP32 throughput to unprecedented levels.\cite{dally2021evolution}

\begin{figure}[H]
    \centering
    \includegraphics[width=\linewidth]{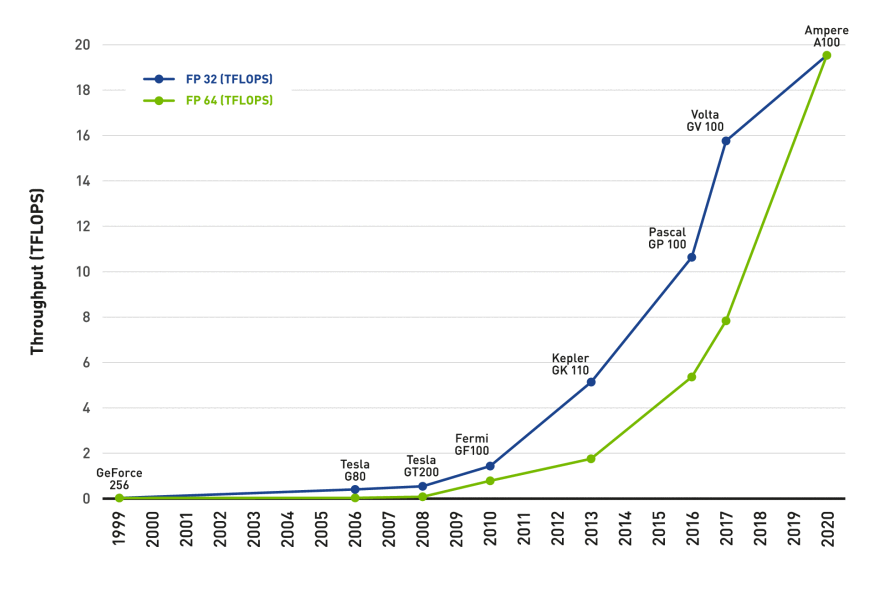} 
    \caption{ Single GPU performance scaling \cite{dally2021evolution}}
    \label{fig:pcg32_vs_kiss99}
\end{figure}

As seen with Huang's Law, which states that GPU performance has been doubling annually since 2012.

Furthermore, GPUs are a consumer-friendly technology, driven by mass-market demand for applications such as VR, 8K, and 16K displays. As these technologies become mainstream, GPUs will become essential parts of personal computing setups, ensuring wide accessibility. This accessibility is crucial for promoting fairness and decentralization in networks like cryptocurrency mining. When every individual can participate with widely available consumer hardware, it mitigates centralization risks posed by specialized hardware like ASICs.

By incorporating FP32 operations, our mining algorithm not only enhances computational efficiency but also provides gaming users with an opportunity to earn rewards without additional investment. This design taps into the vast installed base of consumer GPUs, encouraging decentralized participation and ensuring that the network remains fair and resilient.

\section{Mission and Goals}

PHI's mission is:

\begin{itemize}
    \item \textbf{Decentralization and Fair Participation}: Ensure that every user, whether an ordinary gaming PC user or a small miner, can fairly earn rewards in the PHI network.
    \item \textbf{High Performance}: Provide superior transaction speed and performance to meet the needs of modern financial applications.
    \item \textbf{Becoming the Infrastructure for PoW Cryptocurrencies}: Offer high-performance infrastructure for PoW cryptocurrencies, supporting more applications in the future.
\end{itemize}

\section{Technical Highlights}

\subsection{General Information}

\begin{itemize}
    \item \textbf{Full Name}: PHICOIN
    \item \textbf{Description}: The PoW High-PerformanceInfrastructure
    \item \textbf{Genesis Block}: The Times 11/06/2024: Donald Trump wins US election 2024 to become 47th president.

    \item \textbf{Symbol}: $\Phi$
\end{itemize}

\subsection{Blockchain Specifications}

\begin{itemize}
    \item \textbf{Block Time}: 15 seconds.
    \item \textbf{Block Size}: 4 MB 
    \item \textbf{TPS}: Approximately 1,243 TPS (assuming an average transaction size of 225 bytes)
    \item \textbf{DAG Size}:  > 4 GB
    \item \textbf{DAG Increasing}:  25\% / year 
    \item \textbf{Total Supply}: unlimited
    \item \textbf{Halving}: Yearly
    \item \textbf{Halving Times}: 1
\end{itemize}

\subsection{Mining algorithm requirements}

According to Huang's Law, the performance of GPUs has been increasing at a rate faster than Moore's Law. This trend is evident not only in computational power but also in memory capacity. Over the past two decades, the memory capacity of entry-level graphics cards has grown at an impressive average annual rate of 25.64\%. In 2004, the NVIDIA GeForce 6200 offered only 64MB of memory, while by 2024, the NVIDIA RTX 3050 boasts 6GB, representing a nearly 100-fold increase.

This rapid evolution in GPU memory and performance is crucial for supporting the demands of modern applications, from real-time rendering and machine learning to blockchain operations and cryptographic computations. PHI Coin leverages these advancements to enhance network scalability, security, and efficiency. With GPUs becoming increasingly powerful, PHI Coin ensures its infrastructure remains future-proof by capitalizing on cutting-edge technologies for mining, encryption, and decentralized computing.

\begin{figure}[H]
    \centering
    \includegraphics[width=1\linewidth]{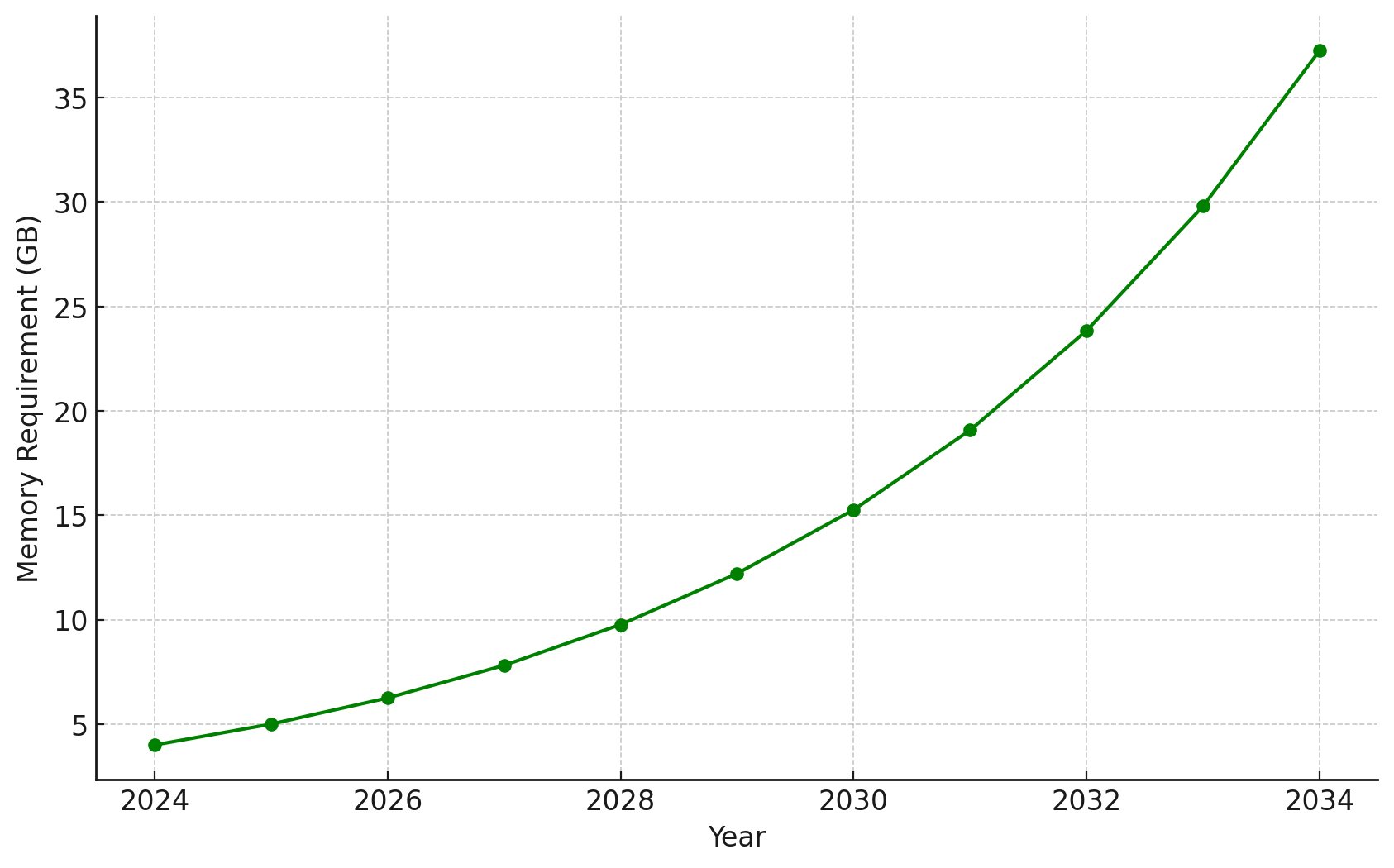} 
    \caption{Evolution of GPU Memory Requirement Over 10 Years}
    \label{fig:pcg32_vs_kiss99}
\end{figure}

The chart above illustrates the planned evolution of GPU memory requirements for PHI coin mining over 10 years. Starting at 4 GB in the first year, the memory requirement increases by 25\% every year, reaching 40 GB by the 10th year. 
\subsection{Mission and Goals}
Unlike Bitcoin’s halving model, PHI Coin adopts an alternative approach by increasing hardware requirements over time. Mining activities are always anchored to the memory capacity of entry-level devices, ensuring that as gamers and consumers naturally upgrade their equipment, they remain eligible to participate. This strategy guarantees fair participation across all timeframes, as everyone can access the network equally using readily available consumer-grade hardware.

We encourage continuous participation while phasing out outdated GPUs as the DAG file grows. By gradually raising the memory threshold, PHI coin ensures that miners remain competitive and profitable while maintaining a sustainable and up-to-date mining ecosystem. This thoughtful upgrade cycle not only extends the lifespan of the mining ecosystem but also supports the ecological balance of the network by discouraging outdated and inefficient hardware.

\begin{figure}[H]
    \centering
    \includegraphics[width=1\linewidth]{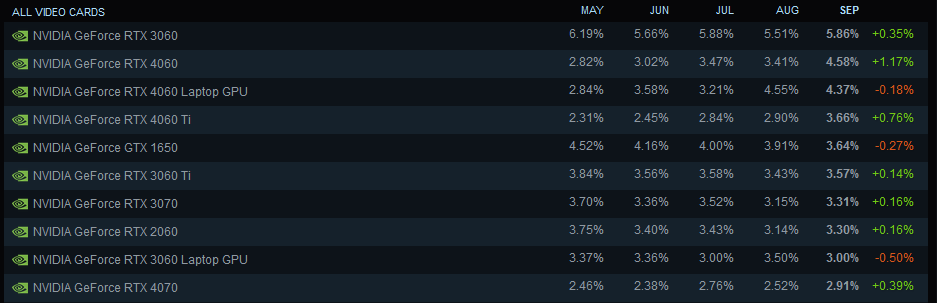} 
    \caption{Steam Hardware Survey: September 2024}
    \label{fig:pcg32_vs_kiss99}
\end{figure}

Based on Steam's data, GPUs with 6GB, 8GB and 12GB of memory are increasingly becoming mainstream, reflecting gamers' growing demand for more powerful hardware capable of handling modern titles and higher resolutions. This shift highlights the importance of keeping mining requirements aligned with these trends, ensuring that participants with popular mid-range and high-end GPUs can continue to mine PHI coin efficiently.\cite{Steam2024}

\begin{figure}[H]
    \centering
    \includegraphics[width=1\linewidth]{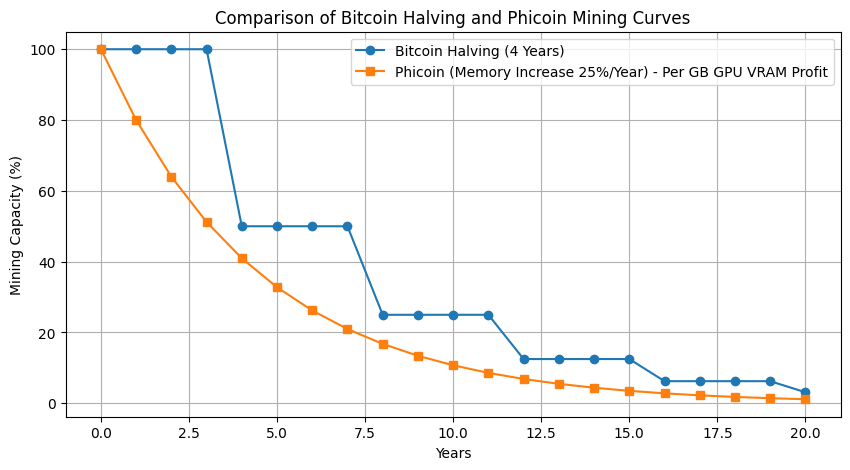} 
    \caption{Comparison of Bitcoin Halving per block and Phicoin Mining Curves per GB vram}
    \label{fig:pcg32_vs_kiss99}
\end{figure}

This chart compares the mining reward models of Bitcoin and Phicoin over 20 years. The Bitcoin Halving Curve shows that the block rewards decrease by half every four years, resulting in a stepped decline in mining capacity. In contrast, the Phicoin Curve demonstrates that with an annual 25\% increase in memory requirements, the mining capacity per GB of GPU VRAM gradually decreases over time.

The Phicoin memory-increase model functions similarly to Bitcoin’s halving mechanism in terms of coin supply control. However, Phicoin’s model has a unique advantage: the rising baseline hardware requirements align with the natural technology upgrade cycle. As people upgrade to more powerful GPUs for gaming, watching high-definition content, or other computational needs, they naturally acquire the required hardware without incurring additional burdens. This process eliminates outdated equipment and ensures fairness within the network by continuously leveling the playing field.

\section{Mining Algorithm: Phihash}

\textbf{Phihash} is an innovative Proof-of-Work (PoW) algorithm designed for the Phicoin blockchain. By integrating key features from Ethash and KawPow/ProgPow, it introduces enhancements to improve security, resist ASIC/FPGA mining hardware, and ensure decentralization. Phihash utilizes FP32 floating-point computations, the Permuted Congruential Generator (PCG) for randomness, and an advanced DAG growth mechanism to achieve fairness and efficiency in GPU mining.

\subsubsection{Key Features of Phihash}

\begin{itemize}
    \item \textbf{ASIC \& FPGA Resistance:} Through randomized memory access and floating-point operations, Phihash minimizes the efficiency of specialized mining hardware.
    \item \textbf{Dynamic DAG Growth:} Phihash introduces a 25\% exponential DAG growth factor per epoch, gradually phasing out outdated GPUs.
    \item \textbf{Optimized GPU Utilization:} FP32 operations maximize GPU parallel computation capabilities, ensuring fairness and high efficiency.
    \item \textbf{Enhanced Randomness and Entropy:} The use of PCG32 ensures high-quality randomness, crucial for mining fairness.
\end{itemize}

\subsubsection{Improved DAG Growth Mechanism}

Phihash implements a dynamic DAG growth strategy inspired by Ethash. Unlike the linear growth in Ethash, Phihash employs an exponential growth factor of 1.25 per epoch, ensuring a gradual transition to newer hardware.

\textbf{Formula for DAG Growth:}
\[
\text{Items per Epoch} = \text{Initial Size} \times (1.25^\text{Epoch Number})
\]

\textbf{Example Calculations:}
\begin{itemize}
    \item \textbf{Initial DAG Size:} 4 GB
    \item \textbf{Epoch 1:} $4 \times 1.25 = 5 \, \text{GB}$
    \item \textbf{Epoch 2:} $5 \times 1.25 = 6.25 \, \text{GB}$
\end{itemize}

\textbf{Code Example:}
\begin{lstlisting}[language=C++, caption={DAG Growth Calculation}]
double growth_factor = 1.25;
int ethash_calculate_full_dataset_num_items(int epoch_number) noexcept {
    double _growth_factor = std::pow(growth_factor, epoch_number);
    int num_items_upper_bound = static_cast<int>(num_items_init * _growth_factor);
    return ethash_find_largest_prime(num_items_upper_bound);
}
\end{lstlisting}

\subsubsection{Randomness with PCG32}

Phihash uses the \textbf{PCG32} pseudo-random number generator to achieve high-quality randomness and enhance resistance to ASIC/FPGA optimizations. PCG32 ensures unpredictable operations in the mining process, making optimization for specialized hardware more difficult.

\textbf{PCG32 Workflow:}
\begin{enumerate}
    \item \textbf{Initialization:} Seed values are derived from the block header hash, nonce, and epoch context.
    \item \textbf{Random Number Generation:} High-entropy 32-bit integers are generated using bitwise shifts and XOR operations.
\end{enumerate}

\textbf{Code Example:}
\begin{lstlisting}[language=C++, caption={PCG32 Implementation}]
class pcg32 {
    uint64_t state;
    uint64_t inc;
public:
    pcg32(uint64_t initstate, uint64_t initseq) : state(initstate), inc(initseq) {}

    uint32_t operator()() {
        uint64_t oldstate = state;
        state = oldstate * 6364136223846793005ULL + (inc | 1);
        uint32_t xorshifted = static_cast<uint32_t>(((oldstate >> 18u) ^ oldstate) >> 27u);
        uint32_t rot = static_cast<uint32_t>(oldstate >> 59u);
        return (xorshifted >> rot) | (xorshifted << ((32 - rot) & 31));
    }
};
\end{lstlisting}

\textbf{Advantages of PCG32:}
\begin{itemize}
    \item High-quality randomness with minimal memory requirements.
    \item Efficient computation on modern GPU hardware.
\end{itemize}

\subsubsection{FP32 Computations for GPU Optimization}

Phihash integrates \textbf{FP32 (single-precision floating-point)} computations to exploit the high parallelism of modern GPUs. This ensures that:
\begin{itemize}
    \item GPU miners achieve maximum efficiency.
    \item Mining remains accessible and fair for ordinary users with modern GPUs.
\end{itemize}

\textbf{Example FP32 Operations:}
\begin{itemize}
    \item Trigonometric functions like $\tanh$, $\sin$, and $\cos$.
    \item Randomized addition, multiplication, and division.
\end{itemize}

These operations increase the computational complexity of the algorithm, reducing the advantage of ASICs and FPGAs.

\subsubsection{PhiHash Algorithm}

The \textbf{PhiHash} algorithm is designed to enhance security and decentralization in the Phicoin blockchain. It consists of several key steps, which are mathematically described as follows:

\begin{enumerate}
    \item \textbf{Initialization of Keccak State:}  
    Initialize the Keccak state using the block header hash and nonce:
    \[
    \text{Keccak\_state} = \text{InitializeKeccak}(\text{header\_hash}, \text{nonce})
    \]

    \item \textbf{Generation of Mix Seed:}  
    Generate a seed for the mix array:
    \[
    \text{Seed} = \text{GenerateSeed}(\text{Keccak\_state})
    \]

    \item \textbf{Initialization of Mix Array:}  
    Initialize the mix array using the generated seed:
    \[
    \text{Mix} = \text{InitializeMix}(\text{Seed})
    \]

    \item \textbf{Mixing Rounds:}  
    Perform a series of mixing rounds to enhance randomness and security. For each round \( i \) from 1 to 64:
    \[
    \text{Mix} = \text{MixRound}(\text{Mix}, \text{epoch\_context})
    \]

    \item \textbf{Finalization of Hash:}  
    Compute the final hash using the Keccak function:
    \[
    \text{Final\_hash} = \text{Keccak}(\text{Mix})
    \]

    \item \textbf{Output:}  
    The final hash is returned as the result of the PhiHash algorithm.
\end{enumerate}

\subsubsection{Security and Decentralization}

Phihash enhances the security and decentralization of the Phicoin blockchain by:
\begin{itemize}
    \item \textbf{Reducing ASIC/FPGA Advantage:} Randomized memory access patterns and FP32 computations reduce the efficiency of specialized mining hardware.
    \item \textbf{Promoting Decentralized Participation:} By requiring GPUs with at least 4 GB of VRAM, Phihash ensures broad participation while phasing out outdated hardware.
\end{itemize}

\textbf{Key Benefits:}
\begin{itemize}
    \item Enhanced network security through high entropy and randomness.
    \item Fair mining opportunities for ordinary GPU users.
\end{itemize}

\subsubsection{Advanced Features of Phihash}

\begin{itemize}
    \item \textbf{Memory Optimization:} Efficient use of DAG-based operations and randomized cache access patterns to utilize GPU L1 and L2 caches.
    \item \textbf{Exponential DAG Growth:} The 25\% growth factor per epoch ensures long-term sustainability while gradually increasing hardware requirements.
\end{itemize}

\subsubsection{Community Mining Tests and Entry-Level Fairness}
Beyond the core design and theoretical underpinnings of Phihash, real-world testing by community members has provided valuable insights into the algorithm’s robustness and accessibility. Various users have run 15-minute mining sessions on their consumer-grade gaming GPUs (often differing in both model and quantity Nvidia GPUs) to verify the stability and fairness of Phihash in practice.

\paragraph{15-Minute Test Results.}
Table~\ref{tab:gpu_hashrate} below summarizes sample hashrates collected from distinct GPU setups, each running the Phihash algorithm for at least 15 minutes:
\begin{itemize}
    \item \emph{Stable Performance:} Despite hardware differences (ranging from older GTX-series models to the latest RTX 4000-series), the hashrates across each 10s, 60s, and 15m window remained relatively consistent, indicating that Phihash adapts well to real-world mining conditions.
    \item \emph{Broad Coverage:} Testing spanned single-GPU laptops, dual-GPU desktops, and multi-GPU mining rigs, demonstrating the algorithm’s versatility.
\end{itemize}

\begin{table*}[!t]
    \centering
    \caption{Hashrates (in MH/s) for Different GPUs over 10s, 60s, and 15m Intervals}
    \label{tab:gpu_hashrate}
    \resizebox{\textwidth}{!}{%
    \begin{tabular}{lccccccc}
    \hline
    \textbf{GPU Model} & \textbf{\# of GPUs} & \textbf{Total (10s) [MH/s]} & \textbf{Total (60s) [MH/s]} & \textbf{Total (15m) [MH/s]} & \textbf{Per GPU (10s) [MH/s]} & \textbf{Per GPU (60s) [MH/s]} & \textbf{Per GPU (15m) [MH/s]} \\
    \hline
    GTX 1060 6GB    & 5 & 45.46 & 44.51 & 40.42 &  9.09 &  8.90 &  8.08 \\
    GTX 1660 SUPER  & 5 & 70.73 & 70.57 & 73.26 & 14.15 & 14.11 & 14.65 \\
    GTX 1080 Ti     & 7 & 145.89 & 154.06 & 155.69 & 20.84 & 22.01 & 22.24 \\
    RTX 2060 SUPER  & 5 & 87.92 & 98.49 & 91.53 & 17.58 & 19.70 & 18.31 \\
    RTX 3060        & 8 & 167.14 & 164.23 & 164.90 & 20.89 & 20.53 & 20.61 \\
    RTX 3070        & 2 & 52.12 & 52.10 & 52.11 & 26.06 & 26.05 & 26.06 \\
    RTX 3080 Laptop & 1 & 26.78 & 26.78 & 26.79 & 26.78 & 26.78 & 26.79 \\
    RTX 3070 Ti     & 2 & 67.91 & 68.66 & 69.75 & 33.96 & 34.33 & 34.88 \\
    RTX 3080 Ti     & 8 & 427.63 & 426.88 & 426.96 & 53.45 & 53.36 & 53.37 \\
    RTX 3090        & 4 & 217.65 & 217.38 & 217.32 & 54.41 & 54.35 & 54.33 \\
    RTX 4060 Ti     & 2 & 34.10 & 34.10 & 34.09 & 17.05 & 17.05 & 17.05 \\
    RTX 4080        & 1 & 42.16 & 48.82 & 42.55 & 42.16 & 48.82 & 42.55 \\
    RTX 4090        & 1 & 59.98 & --    & --    & 59.98 & --    & --    \\
    \hline
    \end{tabular}%
    }
    \end{table*}

\paragraph{Algorithmic Stability and Usability.}
Phihash’s design—from PCG32 randomness to FP32 computations—contributes to a stable hashrate output, even over longer runs. Community feedback highlights:
\begin{itemize}
    \item \textbf{Consistent Difficulty Adjustments:} The network difficulty and DAG size adjustments ensure a smooth mining experience without abrupt drops in efficiency.
    \item \textbf{Gradual Phasing Out of Obsolete GPUs:} As DAG requirements grow, severely outdated or low-VRAM cards naturally become unprofitable, helping maintain a fair playing field.
\end{itemize}

\paragraph{Entry-Level Participation and Decentralization.}
A key principle of Phihash is \emph{inclusivity}: 
\begin{itemize}
    \item \emph{Minimum VRAM Threshold:} With a 4\,GB baseline and progressively rising DAG size, modern entry-level GPUs can comfortably participate. This ensures that hobbyists and small-scale miners have a stake in securing the network.
    \item \emph{Fair Competition:} By leveraging random memory access, FP32 operations, and exponential DAG growth, Phihash discourages centralized advantages that typically arise from ASIC or FPGA deployment.
\end{itemize}

\noindent 
As GPU hardware continues to evolve—especially at the lower end where memory capacity and computational power are rapidly increasing—\textbf{Phihash remains accessible to newcomers}, thus fostering greater decentralization. This design philosophy elevates the role of everyday users, preventing mining monopolies and promoting a more resilient, widely distributed blockchain infrastructure.

\section{Tokenomics}

\subsection{Coin Emission \& Inflation Control}
PHI’s supply strategy ensures that 100\% of the tokens are generated through mining, adhering to the principle of fair distribution. To maintain long-term sustainability and encourage continuous participation, PHI is designed as an unlimited-supply cryptocurrency. However, to control inflation after the second year, we introduce a carefully structured issuance plan.

In the first 14 days, PHI will launch a Testing Mining phase, focused on testing internal blockchain parameters and stability. During this phase, the network will use the Kawpow algorithm to resist ASIC mining, ensuring a fair start for all participants. At the end of this phase, 210{,}240{,}000 PHI will be distributed to all participants based on their contributions during testing.

Once the mainnet goes live, the PHI earned during the Testing Mining phase will be fully allocated to participants. Following the mainnet launch, the issuance schedule is as follows:  
\begin{itemize}
    \item \textbf{First year:} Each block generates 5 PHI.
    \item \textbf{Second year:} Block rewards halve to 2.5 PHI.
    \item \textbf{Subsequent years:} Block rewards stabilize at 2.5 PHI, without further reduction.
\end{itemize}

This issuance plan ensures a controlled inflation rate after the Testing Mining phase. The inflation rates for the first three years are:
\begin{itemize}
    \item 4.76\% after the first year,
    \item 2.32\% after the second year,
    \item 2.22\% after the third year.
\end{itemize}

These inflation levels are carefully calibrated to foster sustainable growth, ensuring the longevity and viability of the PHI project. Block rewards start at 5 PHI per block, halving in the first year, and then remaining fixed at 2.5 PHI per block. This design aims to incentivize early participants by increasing the initial supply and gradually reducing inflation to achieve supply stability, forming a more predictable supply structure.

\begin{figure}[H]
    \centering
    \includegraphics[width=1\linewidth]{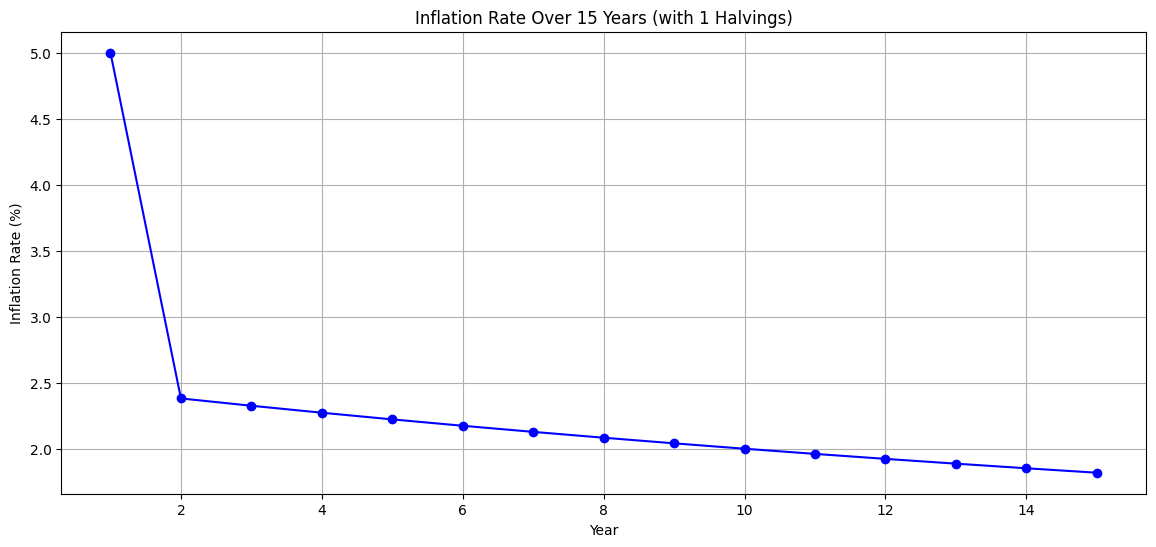}
    \caption{Phicoin's Inflation Rate Over the Next 15 Years}
    \label{fig:pcg32_vs_kiss99}
\end{figure}

By gradually reducing the minting amount and block rewards, PHI provides strong incentives for early adopters while ensuring a low-inflation stable model in the long term, supporting the healthy ecological development of the currency.

\subsection{Coin Distribution}
All PHI tokens are generated exclusively through mining, ensuring that the network remains decentralized and fair. Ninety-five percent (95\%) of the mined tokens are allocated directly to miners, creating a robust incentive structure that rewards participants and ensures fairness.

To support the essential operations of the project, PHI collects a 5\% mining fee, which is allocated as follows:
\begin{itemize}
    \item \textbf{3\%} is dedicated to maintaining core infrastructure, including the operation of fundamental nodes, blockchain explorers, and continuous airdrop activities to foster engagement and participation.
    \item \textbf{2\%} is reserved for developer incentives, encouraging innovation and continuous improvements within the ecosystem.
\end{itemize}

\begin{figure}[H]
    \centering
    \includegraphics[width=1\linewidth]{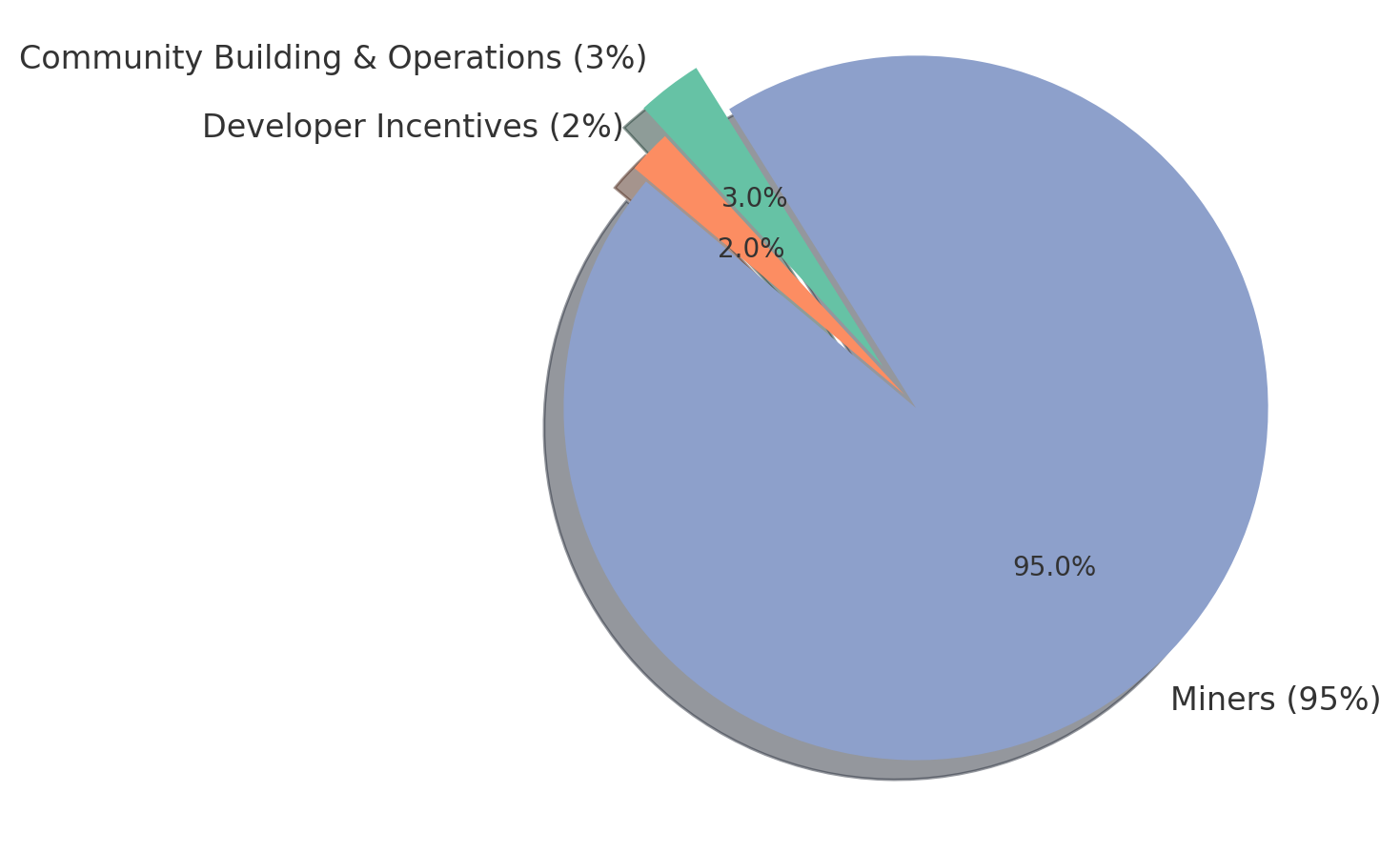}
    \caption{Phicoin Distribution}
    \label{fig:pcg32_vs_kiss99}
\end{figure}

Importantly, these coins are not distributed all at once. Instead, they are gradually released alongside mining activities, ensuring sustainable development and incentivization over the long term. We are committed to fostering an environment of fairness and sustainable growth, ensuring that PHI is a decentralized, participatory project.

We aim to ensure continuous mining opportunities for the community. Even after the second halving, Phicoin's issuance will not converge to zero. As the size of the DAG file increases, older, less efficient GPUs will naturally phase out, allowing miners with newer hardware to remain profitable. Our vision is to create a dynamic ecosystem where innovation and sustainability thrive, ensuring that both early supporters and long-term participants can benefit from mining PHI for years to come.

\begin{figure}[H]
    \centering
    \includegraphics[width=1\linewidth]{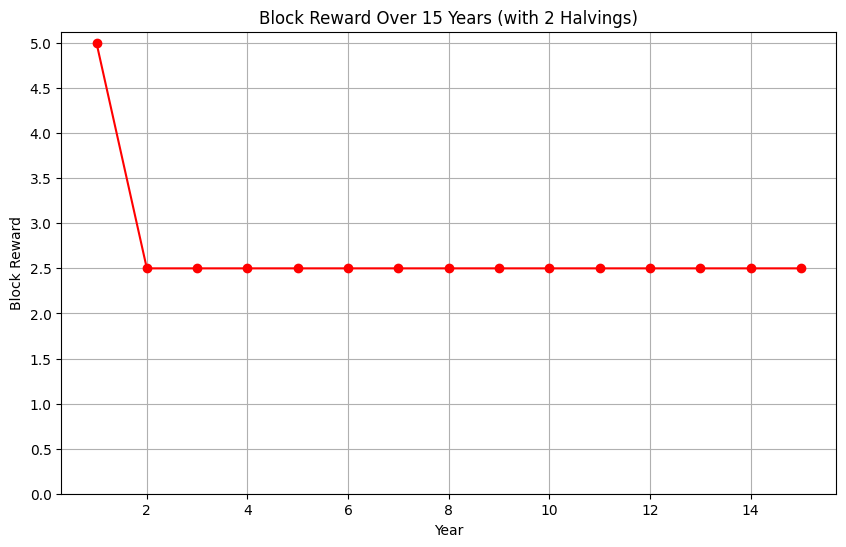}
    \caption{Block rewards for each block over the next 15 years}
    \label{fig:pcg32_vs_kiss99}
\end{figure}

\subsection{Staking Pool}
While traditional Proof-of-Work systems rely primarily on block rewards for miner incentives, Phicoin supplements its PoW model with a dedicated \textit{staking pool}. This mechanism leverages the 5\% mining fee to reward users who stake up to the first 10 million PHI. The goal is to foster long-term holding, reduce token velocity, and enhance overall price stability.

\paragraph{Year 1 Staking Incentives.}
During the first year, each block generates 5 PHI, and 5\% of this reward flows into the staking pool. Let $n_b = 5760$ be the average number of blocks per day and $d = 365$ the number of days per year. If $T = 10{,}000{,}000$ PHI is the total staked amount, then the annual yield $Y_1$ is:
\begin{equation}
\label{eq:stakingYear1}
    Y_1 \;=\; \frac{n_b \times 5 \times d \times 0.05}{T}
    \;=\; 0.05256
    \quad(\!5.256\%\text{~APR}).
\end{equation}
Hence, early participants in the staking pool can receive approximately 5.256\% annualized rewards on their staked PHI.

\paragraph{Year 2 Staking Incentives.}
Starting the second year, the block reward halves to 2.5 PHI. Consequently, the staking fee allocated per block also diminishes. The updated annual yield $Y_2$ becomes:
\begin{equation}
\label{eq:stakingYear2}
    Y_2 \;=\; \frac{n_b \times 2.5 \times d \times 0.05}{T}
    \;=\; 0.02628
    \quad(\!2.628\%\text{~APR}).
\end{equation}

\paragraph{Locking Period and Settlement.}
To strengthen long-term network engagement, the staking pool imposes:
\begin{itemize}
    \item \textbf{Lock-In Duration:} Newly staked tokens must remain locked for 3 months (90 days).
    \item \textbf{Weekly Payouts:} Staking rewards accumulate weekly, providing a regular yield and incentivizing consistent involvement in network activities.
\end{itemize}

\paragraph{Design Rationale and Price Stability.}
By incorporating staking rewards:
\begin{itemize}
    \item \textbf{Lowered Circulating Supply:} Locking PHI in the staking pool decreases short-term availability, curbing price volatility.
    \item \textbf{Community Incentives:} Stakers share in the network’s success through reliable yield, promoting loyalty and active participation.
    \item \textbf{Sustainable Tokenomics:} Staking serves as a complementary mechanism to PoW mining, providing a balanced approach to distribution. Miners secure the network computationally, while stakers stabilize the economy, creating a robust dual-incentive ecosystem.
\end{itemize}

Overall, this staking pool is designed to encourage a stable token price and foster a committed user community, ensuring that \textbf{Phicoin} remains both fair and economically resilient. By merging a traditional PoW structure with a structured staking model, Phicoin enables a broader range of participants to benefit from the network’s growth and long-term sustainability.

\section{Roadmap}

\subsection{Phase 1: Testing Mining Phase}
Launch the Testing Mining pool using the Kawpow algorithm for 14 days to create a fair starting point for all participants. This phase focuses on testing network parameters and stability, ensuring a robust and decentralized foundation for the PHI blockchain.

\subsection{Phase 2: Launch Phase}
Deploy the mainnet along with core infrastructure, including the blockchain explorer and wallets (desktop, web versions). To enhance user accessibility, a web wallet will also be developed, making it easier for users to manage and interact with Phicoin.

\subsection{Phase 3: Phicoin Listings}
Phicoin will seek to list on trading platforms to provide liquidity and value realization for miners' rewards. In this phase, we aim to list Phicoin on at least one small exchange and one mid-sized exchange. This will help miners convert their mining output into tangible value while expanding the accessibility of Phicoin.

\subsection{Phase 4: Establishment of Phi Lab Foundation}
Create a non-profit research institution, the Phi Lab Foundation, dedicated to researching decentralized protocols. This includes developing innovative solutions such as decentralized DNS (DDNS)\cite{ddns} and decentralized AI systems to further expand the use cases and utility of Phicoin.

\subsection{Phase 5: Launching the Staking Pool}
Implement and activate the staking pool on Phicoin’s mainnet. This initiative offers an additional layer of incentives for token holders, rewarding them with a competitive Annual Percentage Yield (APY) for locking their PHI. The staking pool complements PoW mining by:
\begin{itemize}
    \item \emph{Encouraging Long-Term Holding:} Reducing token velocity and stabilizing the network’s overall economy.
    \item \emph{Strengthening Governance:} Allowing stakers to participate more actively in community decisions and future protocol upgrades.
    \item \emph{Enhancing Network Security:} Diversifying the incentive model, balancing mining rewards with staking benefits.
\end{itemize}

\subsection{Phase 6: Development of Decentralized Applications (DApps)}
Develop and deploy decentralized and innovative applications powered by Phicoin. This includes applications like the decentralized DNS protocol (DDNS) to showcase the blockchain’s ability to support next-generation infrastructure for a decentralized internet.

\subsection{Phase 7: Cross-Chain Bridges}
Establish cross-chain bridges between Phicoin and major blockchain networks such as Solana and Ethereum. These bridges will enable Phicoin integration with decentralized exchanges (DEX), enhancing liquidity and use cases across multiple blockchain ecosystems.

\subsection{Phase 8: Decentralized AI Applications and Protocols}
Develop and deploy decentralized AI-powered applications and protocols on the Phicoin blockchain. These solutions will leverage Phicoin’s infrastructure to create cutting-edge, decentralized AI services and protocols, further demonstrating the versatility and scalability of the Phicoin ecosystem.

\subsection{Note on Centralized Exchange (CEX) Listings}

Listing on centralized exchanges (CEX) can be costly, requiring both substantial funds and coins. While we will explore the option of accepting donations to facilitate CEX listings, our primary focus is on building a strong and efficient network. We believe that with a robust user base and well-developed infrastructure, PHI can thrive as a sustainable, evergreen PoW project. It will offer a fair participation option to all cryptocurrency enthusiasts and establish itself as a viable alternative in the blockchain ecosystem.

We hope that PHI can become an excellent PoW choice, encouraging CEXs and market makers to proactively integrate PHI into their platforms.

\section{Current Development and Evaluation}
Building on the roadmap milestones, the Phicoin project has made tangible progress toward its decentralization goals and market adoption. This section outlines key indicators and data points reflecting Phicoin’s recent growth and the effectiveness of its algorithm, economic model, and community engagement.

\subsection{Deployment Status and Community Metrics}

\paragraph{Test Mining Phase.}
During the 14-day test mining period, network statistics indicated a peak of over \textbf{1,800 online miners}, demonstrating robust interest and participation from diverse user groups. This sizable miner turnout validated the ease of entry for GPU miners and offered a large-scale test of Phicoin’s core functionalities.

\begin{figure}[H]
\centering
\includegraphics[width=0.8\linewidth]{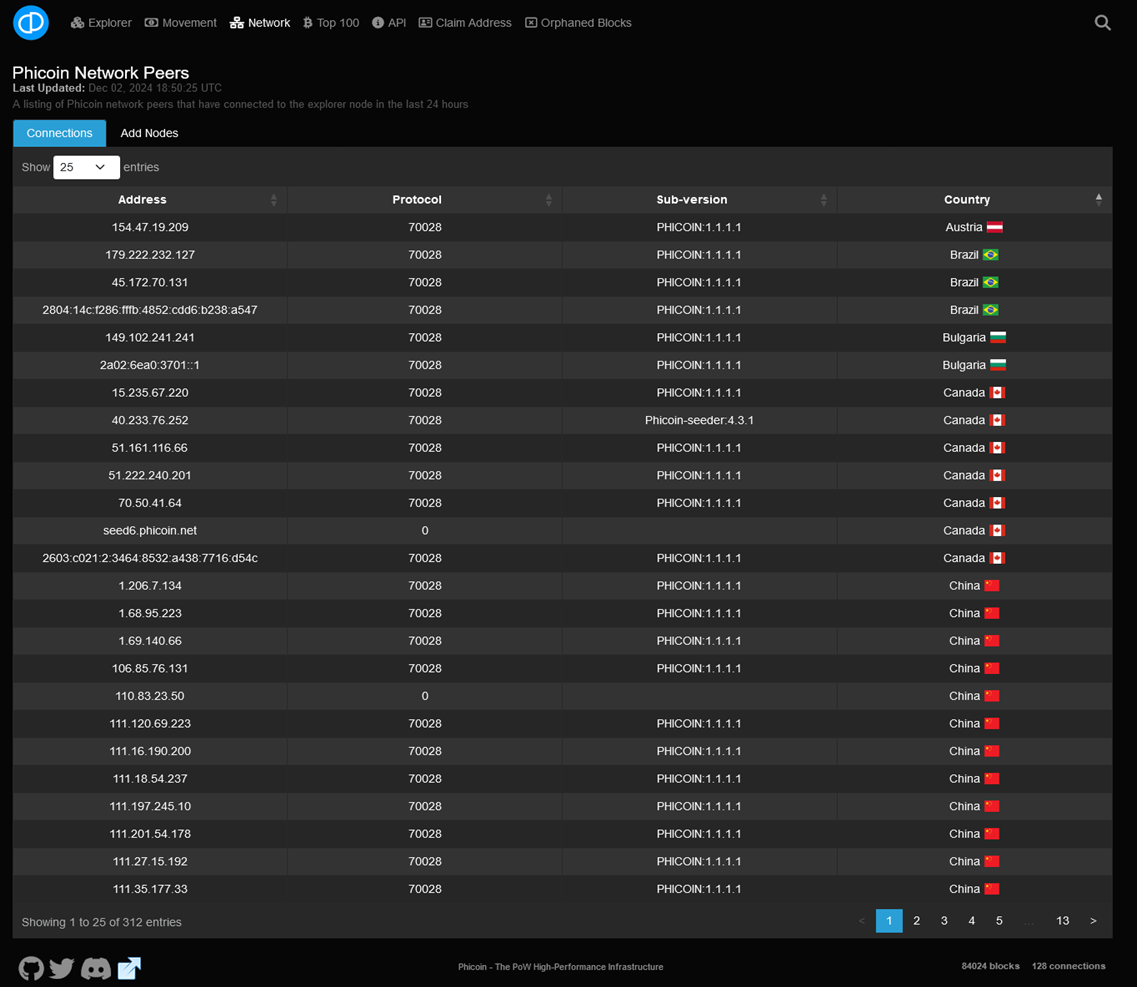} 
\caption{Peak of over 300 active nodes after mainnet launch (Phicoin Explorer)}
\label{fig:peak_nodes}
\end{figure}

\begin{figure}[H]
\centering
\includegraphics[width=1\linewidth]{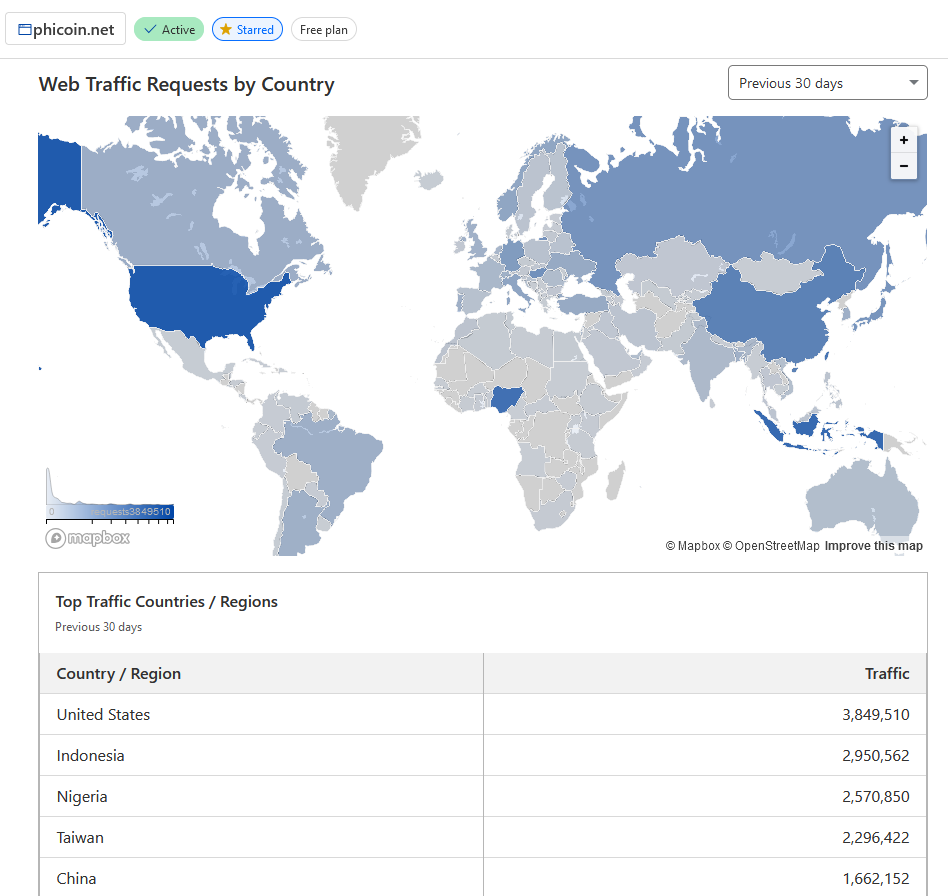} 
\caption{Phicoin node distribution across 10+ countries (Cloudflare)}
\label{fig:geographic_distribution}
\end{figure}

\paragraph{Mainnet Launch and Node Distribution.}
Following the mainnet launch, the network achieved a peak of more than \textbf{300 active nodes} as illustrated in Figure~\ref{fig:peak_nodes}, distributed across at least ten different countries and regions (Figure~\ref{fig:geographic_distribution}). Such geographic dispersion underscores the blockchain’s \emph{decentralized nature}, mitigating the risk of regional concentration and single points of failure.

\begin{figure}[H]
\centering
\includegraphics[width=1\linewidth]{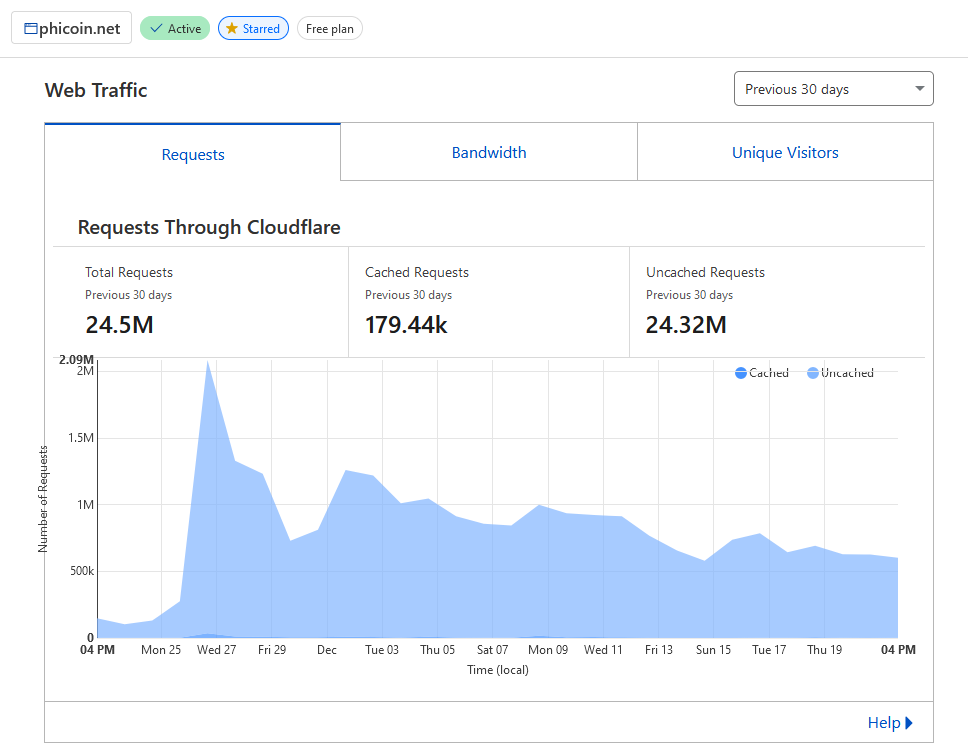} 
\caption{Monthly website requests surpassing 24.5M (Cloudflare)}
\label{fig:traffic}
\end{figure}

\paragraph{Website Traffic.}
Within its first month, the official Phicoin website recorded over \textbf{24.5 million requests} (Figure~\ref{fig:traffic}), highlighting significant inbound interest and organic growth in community awareness. Data from server logs indicates that unique visitor counts showed a healthy upward trend, reinforcing the project’s accessibility and user-friendly design.

\subsection{Market Adoption and Price Behavior}
\begin{figure}[H]
\centering
\includegraphics[width=1\linewidth]{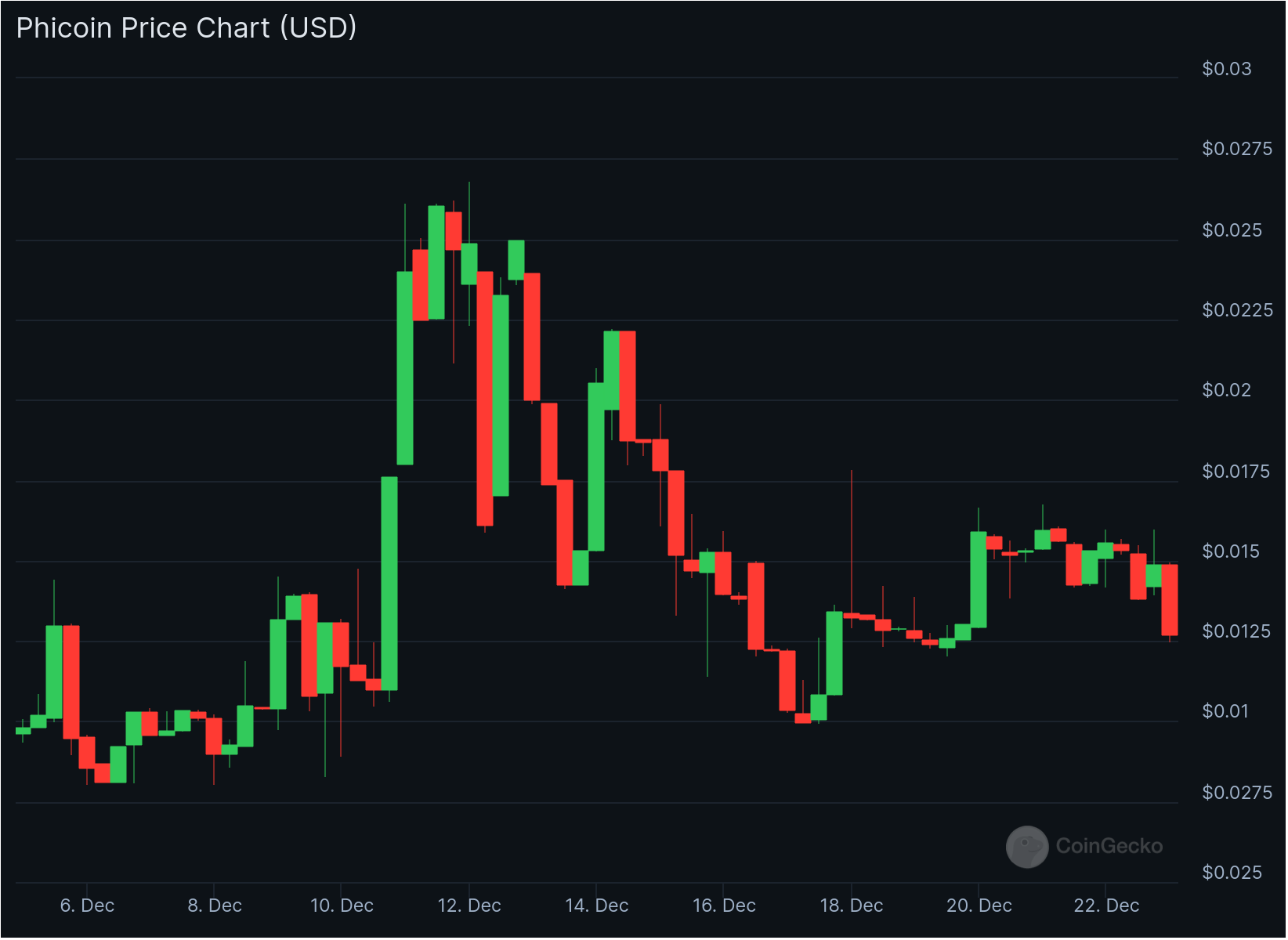} 
\caption{PHI token price movement (CoinMarketCap)}
\label{fig:price_chart}
\end{figure}

Phicoin has successfully listed on the smaller centralized exchange (CEX) \emph{Xeggex}, with an average daily trading volume of about \$20,000 and peak volumes exceeding \$80,000. Initial listing price was \textbf{\$0.01}, and despite selling pressure from early miners, the price stabilized in the range of \textbf{\$0.01--\$0.02} (see Figure~\ref{fig:price_chart}). This stability suggests a balance between mining-based token emission and market demand, reflecting the resilience of Phicoin’s underlying economic model.

\subsection{Empirical Insights and Success Factors}

\paragraph{Algorithmic Viability.}
The relatively high miner participation rate and wide node distribution \emph{empirically confirm} the accessibility of Phihash. The requirement of \(\geq 4\,\mathrm{GB}\) VRAM GPU hardware and the gradual DAG growth mechanism have effectively drawn in hobbyist and small-scale miners without hindering performance or creating an excessive barrier to entry.

\paragraph{Economic Model.}
Stabilized coin price points post-listing suggest that the block reward schedule, inflation control, and staking incentives have collectively facilitated:
\begin{itemize}
    \item \textbf{Orderly Token Distribution:} Large-scale early dumping (“miner sell-off”) did not collapse the market, indicating sustainable demand from both new entrants and existing community members.
    \item \textbf{Network Engagement:} The staking pool provides an alternate avenue to participate, lowering token velocity and encouraging long-term holding without undermining the Proof-of-Work principle.
\end{itemize}

\paragraph{Decentralization.}
Data on miners, node operators, and trading activity strongly imply that no single entity or region dominates the network. With over 300 nodes across ten-plus countries, Phicoin exhibits meaningful \emph{geographical and operational decentralization}, reducing systemic risks and bolstering trust.

\paragraph{Community and Ecosystem Growth.}
A substantial portion of the project’s momentum stems from an active Discord and Twitter following, reportedly exceeding 10,000 users. Their collective contributions—ranging from bug reporting and feature suggestions to hashrate benchmarking—substantially enhance Phicoin’s technical robustness and grassroots credibility.

\subsection{Summary of Real-World Impact}
These empirical findings illustrate how Phicoin’s \emph{PoW algorithm}, \emph{economic design}, and \emph{community-driven approach} jointly foster a dynamic and stable environment. The combination of robust mining participation, steady market liquidity, and distributed node presence underscores the project’s viability and potential for long-term sustainability. Moving forward, the team and community will continue refining the protocol and expanding the ecosystem, leveraging real-world data to guide improvements and maintain a high degree of decentralization.

\section{Future Outlook}

In the short term, we aim for PHI to develop into a GPU-based mining project that offers continuous and sustainable rewards. The name PHI originates from the Greek letter PHI, which symbolizes the golden ratio, a representation of perfection and harmony. PHI is designed to be fair, high-performance, and capable of sustainable generation, ensuring that participants—regardless of their size or resources—can benefit equitably from the network.

In the long term, we envision PHI becoming a foundational infrastructure in the world of Proof-of-Work (PoW) cryptocurrencies. We aspire to expand the meaning and scope of Phicoin beyond its current framework. Through our academic research institution, Phi Lab Foundation, we aim to explore and develop innovative applications such as decentralized domain name resolution protocols (DDNS) and decentralized AI protocols.

In this vision, the "\textbf{I}" in PHI comes to stand for \textbf{I}nfrastructure, \textbf{I}nnovation, and \textbf{I}ntelligence, reflecting our commitment to creating a versatile and forward-thinking ecosystem. PHI is not only a cryptocurrency but also a symbol of future possibilities in decentralized systems and cutting-edge technology.

\section{Acknowledgments}
We gratefully acknowledge the foundational contributions of Bitcoin (BTC), Ethereum (ETH), and Ravencoin (RVN), whose pioneering work in Proof-of-Work (PoW) mining laid the groundwork for GPU-based consensus and inspired further innovation. 

We also extend our deepest thanks to the more than one thousand miners who took part in Phicoin’s test mining phase, dedicating their hardware and time to refine the network. The enthusiastic support of our community has been instrumental in shaping a more secure and efficient ecosystem. In particular, we wish to thank the Discord user \texttt{@wapisnet} for conducting comprehensive Phihash tests on a wide range of mainstream Nvidia GPUs, providing invaluable data and insights.

\end{document}